\documentclass[aps,prd,nofootinbib,12pt]{revtex4}
    \usepackage{graphicx}
    \usepackage{amsmath,amssymb,mathrsfs}
\usepackage{epsf}
\usepackage{slashed}
\usepackage{epsfig}
\usepackage{xcolor}
\allowdisplaybreaks

\newcounter{fig}   \newcommand{\lbfig}[1]{\refstepcounter{fig}
\label{#1} }

\newcommand{\bea}{\begin{eqnarray}}
\newcommand{\eea}{\end{eqnarray}}
\newcommand{\be}{\begin{equation}}
\newcommand{\ee}{\end{equation}}

\newcommand{\re}[1]{(\ref{#1})}

\newcommand{\eqn}{\begin{eqnarray}}
\newcommand{\eqnx}{\end{eqnarray}}

\begin{document}

\title{Kerr black holes with parity-odd scalar hair}

\author{J.~Kunz}
\affiliation{Institute of Physics, University of Oldenburg, Germany
Oldenburg D-26111, Germany
}
\author{I.~Perapechka}
\affiliation{
Department of Theoretical Physics and Astrophysics,
Belarusian State University, Minsk 220004, Belarus}
\author{Ya.~Shnir}
\affiliation{BLTP, JINR, Dubna 141980, Moscow Region, Russia\\
Department of Theoretical Physics, Tomsk State Pedagogical University, Russia
}

\begin{abstract}
We study Kerr black holes with synchronised non-trivial
parity-odd massive scalar hair
in four dimensional asymptotically flat space-time.
These axially symmetric stationary spinning solutions of the
minimally coupled Einstein-Klein-Gordon theory
provide yet another example of bound states
in synchronous rotation with the event horizon.
We discuss the properties of these parity-odd hairy black holes and boson stars
and exhibit their domain of existence.
Considering the ergo-regions of these hairy black holes, we show that apart from
the previously discussed ergo-sphere and ergo-Saturn, they
support a new type of composite ergo-surfaces with the topology of
a double-torus-Saturn $(S^1\times S^1)\bigoplus(S^1\times S^1)\bigoplus S^2$.
\end{abstract}
\maketitle

\section{Introduction}
One of the interesting recent developments in  General Relativity
is related to the discovery of new families of stationary, asymptotically  flat
black  holes (BHs), which circumvent the well-known ``no-hair'' theorem
(see e.g.~\cite{Herdeiro:2015waa,Volkov:2016ehx}  and references therein).
An interesting type of such hairy BHs emerge in Einstein-Klein-Gordon theory
with a massive complex scalar field, 
representing stationary spinning BHs with synchronised hair
\cite{Hod:2012px,Herdeiro:2014goa,Herdeiro:2014jaa,Herdeiro:2015gia}.
The key ingredient in the construction of these solutions
is the synchronization condition between the angular velocity of the event horizon
and a phase frequency of the scalar field,
which secures the absence of scalar flux through the event horizon
\cite{Hod:2012px,Herdeiro:2014goa,Herdeiro:2014jaa,Herdeiro:2015gia}.
Many related solutions, relying on this synchronisation mechanism, have been found in the
last few years,
see e.g.~\cite{Hod:2014baa,Benone:2014ssa,Herdeiro:2014pka,Herdeiro:2014ima,Kleihaus:2015iea,Herdeiro:2015tia,Herdeiro:2015kha,Herdeiro:2016tmi,Brihaye:2016vkv,Hod:2017kpt,Herdeiro:2017oyt,Herdeiro:2018daq,Herdeiro:2018djx,Wang:2018xhw,Delgado:2019prc}.

These rotating hairy BHs can be dynamically linked to the Kerr solution
via the superradiant instability mechanism \cite{East:2017ovw,Herdeiro:2017phl}.
On the other hand, in the limit of vanishing event horizon,
they reduce to the globally regular rotating self-gravitating boson stars
\cite{Schunck:1996he,Ryan:1996nk,Yoshida:1997qf,Kleihaus:2005me,Kleihaus:2007vk}.
In the limit where the scalar field trivialises,
these rotating hairy BHs reduce to the usual vacuum Kerr BHs in General Relativity.
Further, they may possess interesting novel physical features. For instance,
they may possess disjunct ergo-regions
consisting of an ergo-sphere surrounded by an ergo-torus
(i.e., a so-called ``ergo-Saturn'') \cite{Herdeiro:2014jaa,Kleihaus:2015iea}.

The family of rotating hairy BHs with synchronised hair
is characterized both by the usual set of physical quantities
like mass, angular momentum and Noether charge,
as well as by two integers: the azimuthal winding number of the complex scalar
field $n$, often referred to as its rotational quantum number,
and its node number $k$.
Most of the studies have focused on the nodeless fundamental solutions
with $n=1$ and $k=0$.
Only recently also radially and angularly excited hairy BHs
with $k\neq 0$ \cite{Wang:2018xhw},
and rotationally excited hairy BHs with $n > 1$
were considered in \cite{Delgado:2019prc}
(see also the case of scalarized hairy BHs \cite{Kleihaus:2015iea}).

By analogy with axially-symmetric spinning Q-balls in flat space
\cite{Volkov:2002aj,Kleihaus:2005me,Kleihaus:2007vk,Radu:2008pp,Kunz:2013wka,Loiko:2018mhb},
as well as based the existence of the corresponding
perturbative Q-cloud solutions
\cite{Benone:2014ssa,Herdeiro:2014pka},
for each value of integer winding number $n$,
there should be two types of spinning solutions possessing
different parity, so called parity-even and parity-odd rotating hairy BHs.

So far, however, only parity-even rotating hairy BHs were investigated
in detail, whereas the parity-odd rotating hairy BHs received little
attention \cite{Wang:2018xhw}.
In this letter we construct a large set of
axially symmetric stationary parity-odd hairy BHs
in the massive complex Klein-Gordon theory
minimally coupled to Einstein's gravity
and discuss their physical properties.

We show that a distinctive new feature of the parity-odd hairy BH solutions
is related to the shapes of their ergo-regions, which,
may represent an ergo-sphere,  an ergo-Saturn, or
a new type of composite ergo-surface,
representing an ergo-double-torus-Saturn,
where an ergo-sphere is surrounded by two ergo-tori located symmetrically
with respect to the equatorial plane.

\section{The model}
We consider the Einstein-Klein-Gordon model,
that is the theory of a single non self-interacting
massive complex scalar field $\phi$, which is  minimally
coupled to Einstein gravity in an asymptotically flat 3+1 dimensional space-time.
The action of the model is
\be
\label{action}
S = \int{\sqrt{-g}\left(\frac{R}{4 \alpha^2}-\mathcal{L}_\mathrm{m}\right) d^4 x},
\ee
where $R$ is the Ricci scalar curvature with respect to the Einstein metric $g_{\mu\nu}$,
$g$ is the determinant of the metric tensor, $\alpha^2=4\pi G$ is the gravitational
coupling constant with Newton's constant $G$,
and $\mathcal{L}_\mathrm{m}$ is the matter field Lagrangian
\be
\label{lagFLS}
\mathcal{L}_\mathrm{m} = \left|\partial_\mu\phi\right|^2 + \mu^2|\phi|^2,
\ee
where $\mu$ is the mass of scalar field.

The action \re{action} is invariant with respect to global $\mathrm{U}(1)$ transformations
of the complex scalar field, $\phi\to\phi e^{i\chi }$, where $\chi$ is a constant.
Associated with this symmetry is the Noether 4-current
\be
\label{Noether}
j_\mu = i(\phi\partial_\mu\phi^\ast-\phi^\ast\partial_\mu\phi)\, ,
\ee
and the corresponding conserved charge is $Q=\int{\sqrt{-g}j^t d^3 x}$.

Variation of the action \re{action} with respect to the metric
leads to the Einstein equations
\be
\label{Einstein}
R_{\mu\nu}-\frac{1}{2}Rg_{\mu\nu}=2\alpha^2 T_{\mu\nu},
\ee
where
\be
\label{SET}
T_{\mu\nu}=\left(\partial_{\mu}\phi\partial_{\nu}\phi^\ast
+\partial_{\nu}\phi\partial_{\mu}\phi^\ast\right) -\mathcal{L}_\mathrm{m}g_{\mu\nu}\, ,
\ee
is the stress-energy tensor of the scalar field.

The corresponding equation of motion for the scalar field
is the linear Klein-Gordon equation
\be
\label{scaleq}
    \left(\Box -\mu^2\right)\phi=0\, ,
\ee
where $\Box$ represents the covariant d'Alembert operator.
It should be noted that, both parameters of the model \re{action},
$\alpha$ and $\mu$, can be rescaled away
via transformations of the coordinates and the field,
$x^\mu\to\frac{x^\mu}{\mu}$ and $\phi\to\frac{\phi}{\alpha}$.
In the numerical calculations we fix these parameters to $\alpha=0.5$ and $\mu=1$.

\subsection{Spinning axially-symmetric configurations}
To obtain stationary spinning axially-symmetric field configurations,
we take into account the presence of two commuting
Killing vector fields $\xi=\partial_t$ and $\eta=\partial_\varphi$,
where  $t$ and $\varphi$ are
the time and azimuthal coordinates, respectively. In these coordinates
the metric can be written in Lewis-Papapetrou form
\be
\label{metrans}
ds^2=-F_0 dt^2 +F_1\left(dr^2+r^2 d\theta^2\right)+ r^2\sin^2 \theta F_2  \left(d\varphi-\frac{W}{r} dt\right)^2,
\ee
where the four metric functions $F_0, F_1, F_2$ and $W$ depend on $r$ and $\theta$ only.

The axially-symmetric ansatz for the stationary scalar field is
\be
\label{scalans}
\phi=f(r,\theta)e^{i(\omega t+n\varphi)},
\ee
where $\omega$ is the frequency of field,
and $n\in\mathbb{Z}$ is the azimuthal winding number
\footnote{Note that we are using the rescaled dimensionless frequency
$\omega \to \omega/\mu$ in the numerical calculations.}.

By substituting the ansatz \re{scalans} into the field equation \re{scaleq}, we obtain
\be
\label{KG-par}
\frac{1}{\sqrt{-g}}\frac{\partial}{\partial r}\left(g^{rr}\sqrt{-g}\frac{\partial f}{\partial r} \right)
+
\frac{1}{\sqrt{-g}}\frac{\partial}{\partial \theta}\left(g^{\theta\theta}\sqrt{-g}\frac{\partial f}
{\partial \theta} \right) - \left( n^2 g^{\varphi \varphi}-2g^{\varphi t} +\omega^2 g^{tt}\right) f
= \mu^2 f \, .
\ee
This implies in the case of the Kerr background,
that the variables separate, $f(r,\theta) = R_{ln}(r) S_{ln}(\theta)$,
where $S_{ln}$ are the spheroidal harmonics, $-l\le n \le l$
and $R_{ln}(r)$ the radial functions \cite{Hod:2012px,Herdeiro:2015waa}.
A linear analysis of the spatial asymptotic behavior of the field,
where the metric functions approach the flat space limit,
shows that the scalar field is proportional to $e^{-\sqrt{\mu^2-\omega^2}\, r }$,
thus for $\omega < \mu$ the scalar field is exponentially localized.

The structure of the dynamical equation \re{KG-par} suggests that,
analogous to the case of the spinning axially-symmetric Q-balls
and boson stars
\cite{Volkov:2002aj,Kleihaus:2005me,Kleihaus:2007vk,Radu:2008pp,Kunz:2013wka,Loiko:2018mhb},
the hairy BH solutions of the Einstein-Klein-Gordon system
may also be either symmetric with respect to reflections in the equatorial
plane, $\theta \to \pi -\theta$, or antisymmetric,
as conjectured in \cite{Herdeiro:2014pka}.
The solutions of the first type are referred to as parity-even,
while the configurations of the second type are termed parity-odd.
While previous studies of rotating hairy BHs focused entirely on the case
of parity-even solutions, see e.g.
\cite{Hod:2014baa,Benone:2014ssa,Herdeiro:2014pka,Herdeiro:2014ima,Kleihaus:2015iea,Herdeiro:2015tia,Herdeiro:2015kha,Herdeiro:2016tmi,Brihaye:2016vkv,Hod:2017kpt,Herdeiro:2017oyt,Herdeiro:2018daq,Herdeiro:2018djx,Wang:2018xhw,Delgado:2019prc},
in this paper we shall construct and discuss parity-odd solutions.

We assume the existence of a rotating event horizon,
located at a constant value of radial variable $r=r_h>0$.
The Killing vector of the horizon is the helicoidal vector field
\be
\label{Killingrh}
\chi=\xi +\Omega_h \eta\, ,
\ee
where the horizon angular velocity is fixed by the value of the metric function $W$
on the horizon
$$
\Omega_h =  -\frac{g_{\phi t}}{g_{tt}}\biggl.\biggr|_{r=r_h} = W\biggl.\biggr|_{r=r_h} \, .
$$
The presence of a rotating horizon allows for the formation of stationary scalar clouds,
supported by the synchronisation condition
\cite{Hod:2012px,Herdeiro:2014goa,Herdeiro:2014jaa,Herdeiro:2015gia}
\be
\label{synchron}
\omega = n \Omega_h \,
\ee
between the event horizon angular velocity $\Omega_h$,
the scalar field frequency $\omega$, and the winding number $n$.
This condition implies that there is no flux of scalar field into the black hole
\cite{Hod:2012px,Herdeiro:2014goa,Herdeiro:2014jaa,Herdeiro:2015gia}.

It is convenient to make use of the exponential parametrization of the metric fields
\be
\label{subst}
F_0=\frac{\left(1-\frac{r_h}{r}\right)^2}{\left(1+\frac{r_h}{r}\right)^2}e^{f_0},
\quad F_1=\left(1+\frac{r_h}{r}\right)^4 e^{f_1}
\quad F_2=\left(1+\frac{r_h}{r}\right)^4 e^{f_2}\, .
\ee
Then a power series expansion near the horizon yields the conditions
of regularity of the profile function $f$ and the metric functions $f_i$
\be
\label{bchor}
\partial_r f \bigl.\bigr|_{r=r_h}=\partial_r f_0\bigl.\bigr|_{r=r_h}=\partial_r f_1\bigl.\bigr|_{r=r_h}
=\partial_r f_2\bigl.\bigr|_{r=r_h} = 0\, .
\ee
These boundary conditions supplement the synchronization condition \re{synchron}
imposed on the metric function $W$.

The requirement of asymptotic flatness  at spatial infinity yields
\be
\label{bcinf}
f_0\bigl.\bigr|_{r\to \infty}
=f_1\bigl.\bigr|_{r\to \infty}=f_2\bigl.\bigr|_{r\to \infty}
=W\bigl.\bigr|_{r\to \infty}=0, \quad f \bigl.\bigr|_{r\to \infty}=0\,  .
\ee
Axial symmetry and regularity impose the following boundary conditions
on the symmetry axis at $\theta=0,\pi$
\be
\label{bcpole}
f\bigl.\bigr|_{\theta = 0,\pi} =
\partial_\theta f_0\bigl.\bigr|_{\theta = 0,\pi} =
\partial_\theta f_1\bigl.\bigr|_{\theta = 0,\pi} =
\partial_\theta f_2\bigl.\bigr|_{\theta = 0,\pi} =
\partial_\theta W\bigl.\bigr|_{\theta = 0,\pi}=0 \, .
\ee
The symmetry of the solutions with respect to the reflections in
the equatorial plane allows us to consider the range of values
of the angular variable $\theta \in [0,\pi/2]$.
The corresponding boundary conditions are
\be
\label{bcaxis}
\partial_\theta f_0\bigl.\bigr|_{\theta = \frac{\pi}{2}} =
\partial_\theta f_1\bigl.\bigr|_{\theta=\frac{\pi}{2}}
= \partial_\theta f_2\bigl.\bigr|_{\theta=\frac{\pi}{2}}
= \partial_\theta W\bigl.\bigr|_{\theta=\frac{\pi}{2}}=0 \, ,
\ee
and for the parity-even solutions
$\partial_\theta f\bigl.\bigr|_{\theta = \frac{\pi}{2}} = 0$,
while for parity-odd solutions $f\bigl.\bigr|_{\theta = \frac{\pi}{2}} = 0$.

Note that the condition of the absence of a conical singularity
requires that the deficit angle should vanish,
$\delta=2\pi\left(1 - \lim\limits_{\theta\to 0} \frac{F_2}{F_1}\right) = 0$.
Hence the solutions should satisfy the constraint
$F_2\bigl.\bigr|_{\theta = 0}=F_1\bigl.\bigr|_{\theta = 0}$.
In our numerical scheme we explicitly
verified (within the numerical accuracy) this condition on the symmetry axis.

Asymptotic expansions of the metric functions at the horizon
and at spatial infinity yield important physical properties of the BHs.
The total ADM mass $M$ and the angular momentum $J$ of the spinning hairy BHs
can be read off from the asymptotic subleading behaviour
of the metric functions as $r\to \infty$
\be
\label{ADM}
g_{tt}=-1 + \frac{\alpha^2 M}{\pi r}+O\left(\frac{1}{r^2}\right), \quad
g_{\varphi t}=\frac{\alpha^2 J}{\pi r}\sin^2 \theta+O\left(\frac{1}{r^2}\right).
\ee

The ADM charges can be represented as sums of the contributions
from the event horizon and the scalar hair:
$M = M_h+M_\Phi$ and $J = J_h+J_\Phi$, respectively. These contributions
can be evaluated separately from Komar integrals
\be
\label{Komar}
\begin{split}
    M_h&=-\frac{1}{2\alpha^2}\oint_\mathrm{S}{dS_{\mu\nu} \nabla^\mu \xi^\nu},
    \quad J_h=\frac{1}{4\alpha^2}\oint_\mathrm{S}{dS_{\mu\nu} \nabla^\mu \eta^\nu},\\
    M_\Phi&=-\frac{1}{\alpha^2}\int_{V}{dS_{\mu} \left(2 T^\mu_\nu \xi^\nu - T\xi^\mu\right)},
    \quad J_\Phi=\frac{1}{2\alpha^2}\int_\mathrm{V}{dS_{\mu} \left(T^\mu_\nu \eta^\nu - \frac12 T\eta^\mu\right)},
\end{split}
\ee
where $S$ is the horizon 2-sphere and $V$ denotes
an asymptotically flat spacelike hypersurface bounded by the horizon.

Note that, analogous to the angular momentum of the stationary rotating boson stars,
there is the quantization relation for the angular momentum of the scalar field,
$J_\Phi=n Q$, where $Q$ is the Noether charge
and $n$ is the winding number of the scalar field.

The horizon properties include the Hawking temperature $T_h$,
which is is proportional to the surface gravity
$\kappa^2=-\frac{1}{2}\nabla_\mu\chi_\nu\nabla^\mu\chi^\nu$,
\be
\label{horQ}
T_h=\frac{\kappa}{2\pi} = \frac{1}{16\pi r_h}\exp\left[\left(f_0 - f_1\right)\bigl.\bigr|_{r = r_h}\right]\, ,
\ee
where $\chi$ is the horizon Killing vector \re{Killingrh}.
Another property of considerable interest is the horizon area $A_h$, given by
\be
\label{horQ-A}
A_h = 32\pi r_h^2 \int_0^\pi d\theta \sin\theta\exp\left[\left(f_1 + f_2\right)\bigl.\bigr|_{r = r_h}\right] \, .
\ee
The rotating hairy BHs satisfy the Smarr relation
\cite{Herdeiro:2014goa,Herdeiro:2014jaa,Herdeiro:2015gia}
\be
\label{Smarr}
M = 2T_h S + 2\Omega_h J_h + M_\Phi,
\ee
where $S=\frac{\pi}{\alpha^2}A_h$ is the entropy of the black hole
and $M_\Phi$ is the scalar field energy outside the event horizon \re{Komar}.
The hairy black holes also satisfy the first law of thermodynamics
\cite{Herdeiro:2014goa,Herdeiro:2014jaa,Herdeiro:2015gia}
$$
dM = T_h dS + \Omega_h d J \, .
$$

\section{Numerical results}

\begin{figure}[p!]
    \begin{center}
        \includegraphics[width=.58\textwidth, trim = 40 20 90 20, clip = true]{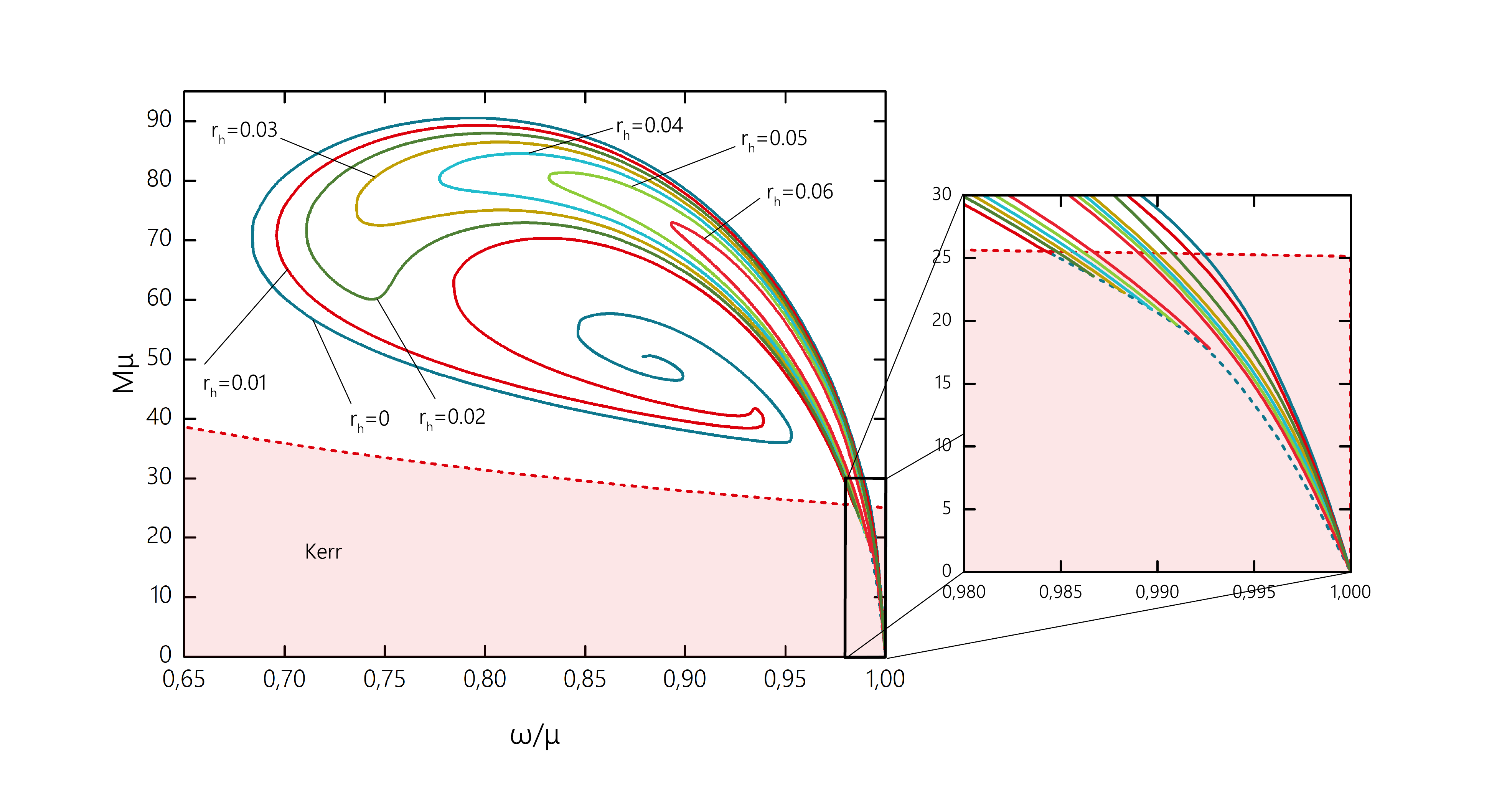}
        \includegraphics[width=.4\textwidth, trim = 40 20 90 20, clip = true]{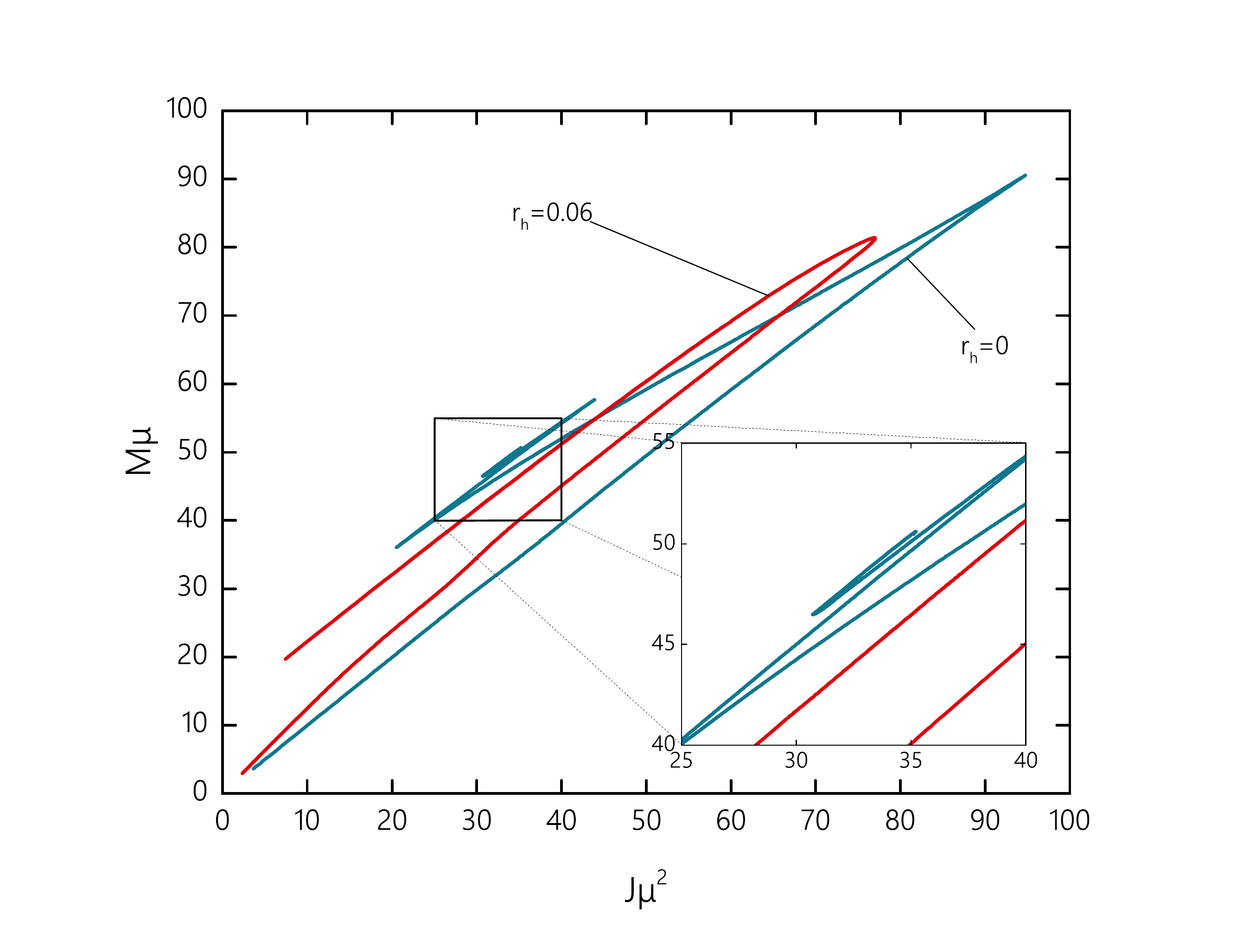}
        \includegraphics[width=.48\textwidth, trim = 40 20 90 20, clip = true]{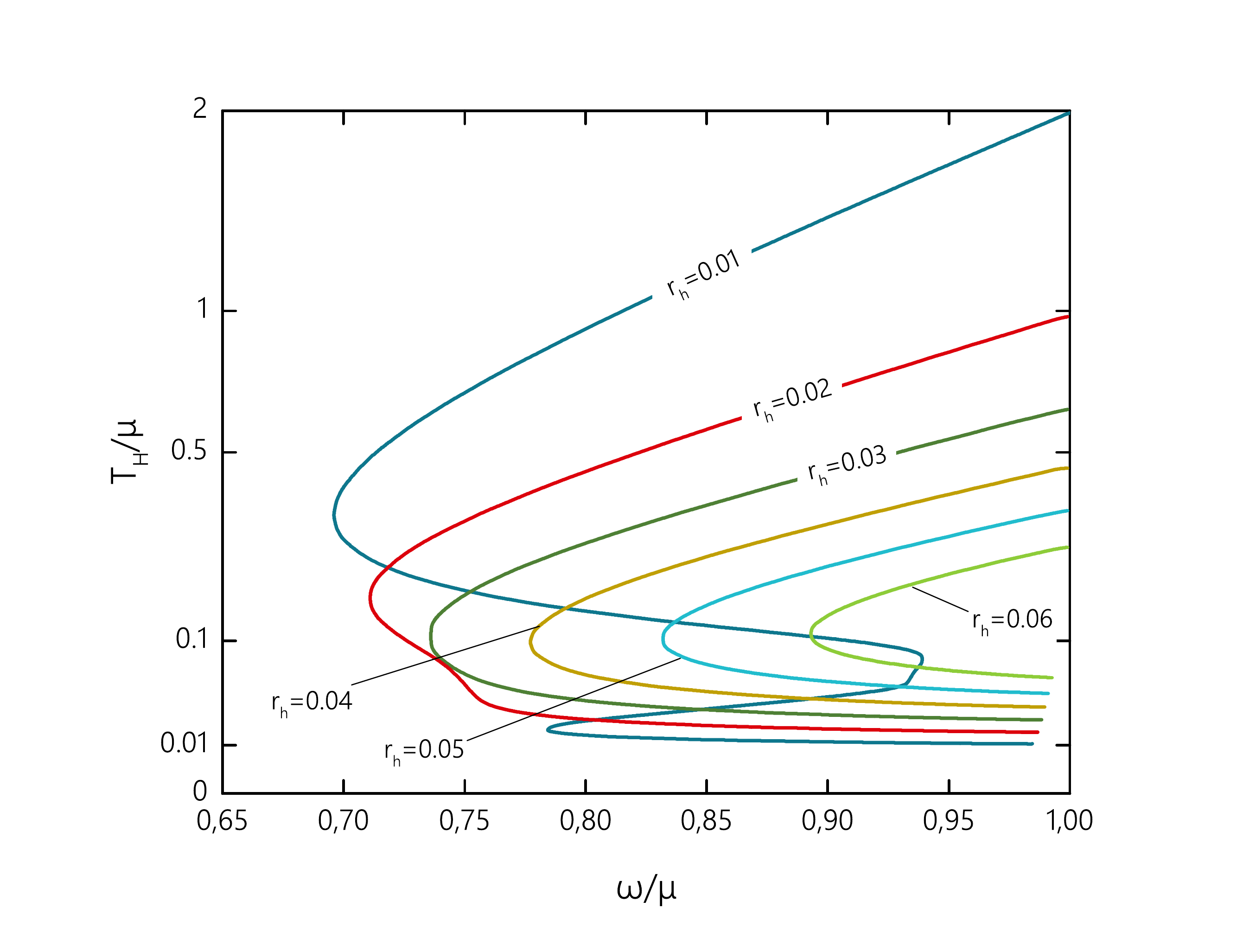}
        \includegraphics[width=.48\textwidth, trim = 40 20 90 20, clip = true]{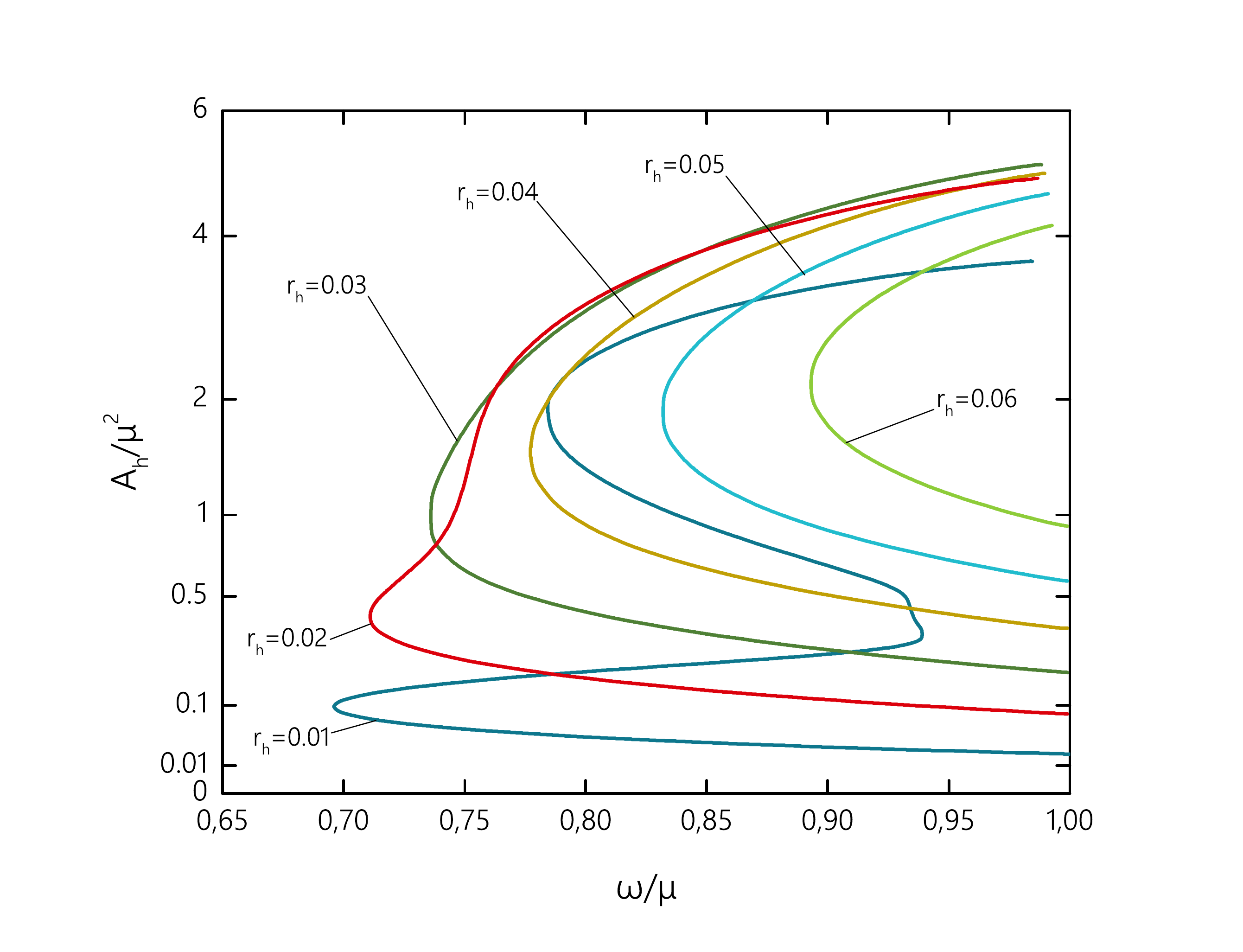}
        \includegraphics[width=.48\textwidth, trim = 40 20 90 20, clip = true]{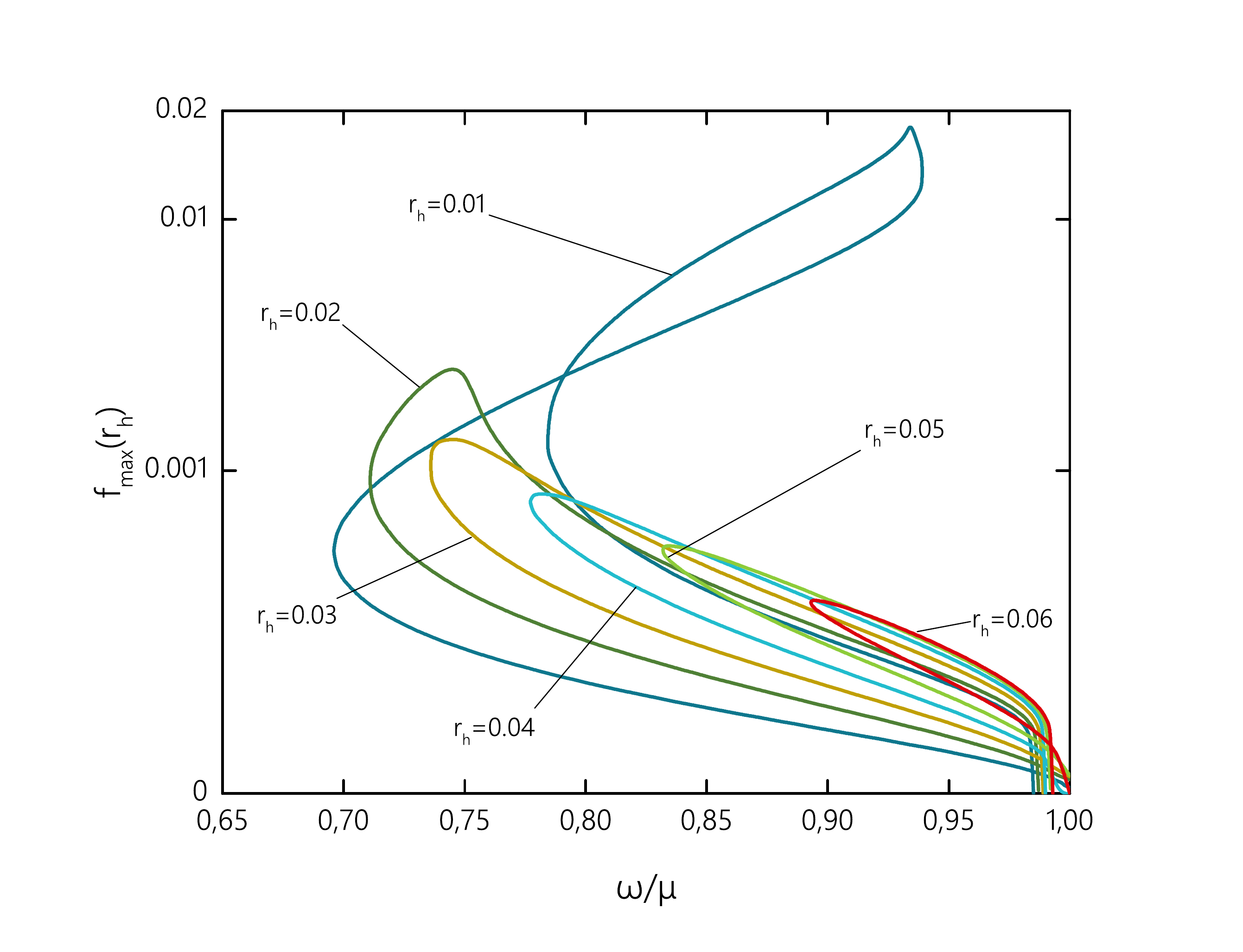}
        \includegraphics[width=.48\textwidth, trim = 40 20 90 20, clip = true]{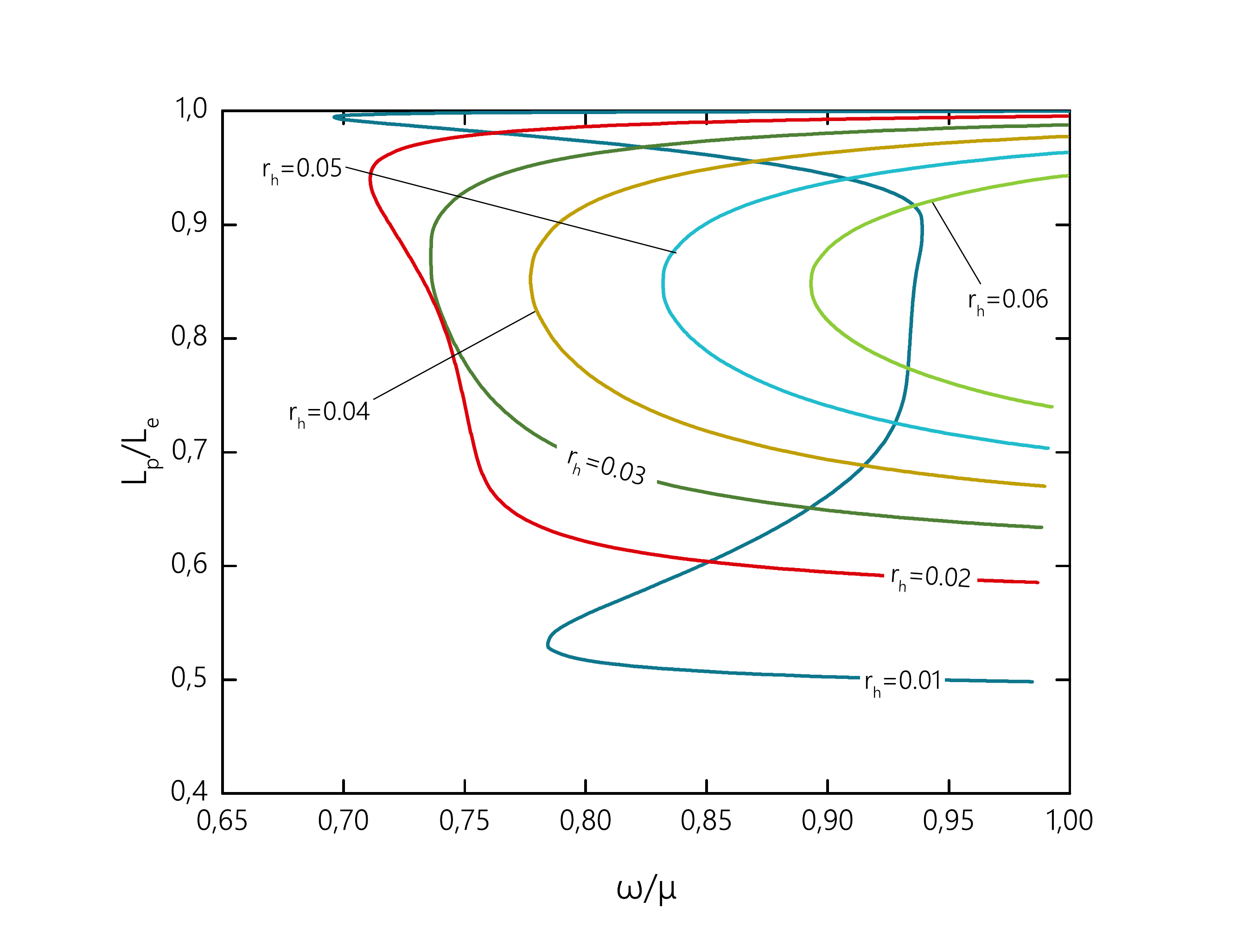}
    \end{center}
    \caption{\small
Properties of parity-odd $n=1$ boson stars and hairy BHs:
ADM mass $M$ vs frequency $\omega$,
where the shaded area corresponds to the domain of existence of vacuum Kerr BHs (upper left),
ADM mass $M$ vs angular momentum $J$ (upper right),
Hawking temperature $T_h$ vs frequency $\omega$ (middle left),
horizon area $A_h$ vs frequency $\omega$ (middle right),
maximal value of the scalar field on the event horizon
$f_\mathrm{max}(r_h)$ vs frequency $\omega$ (lower left),
and the horizon deformation $\epsilon= \frac{L_\mathrm{p}}{L_\mathrm{e}}$
in terms of the ratio of the polar and equatorial circumferences
vs frequency $\omega$ (lower right),
for a set of values of the horizon radius parameter $r_h$.}
    \lbfig{omega_rh}
\end{figure}

\begin{figure}[hbt]
    \begin{center}
        \includegraphics[width=.7\textwidth, trim = 40 20 90 20, clip = true]{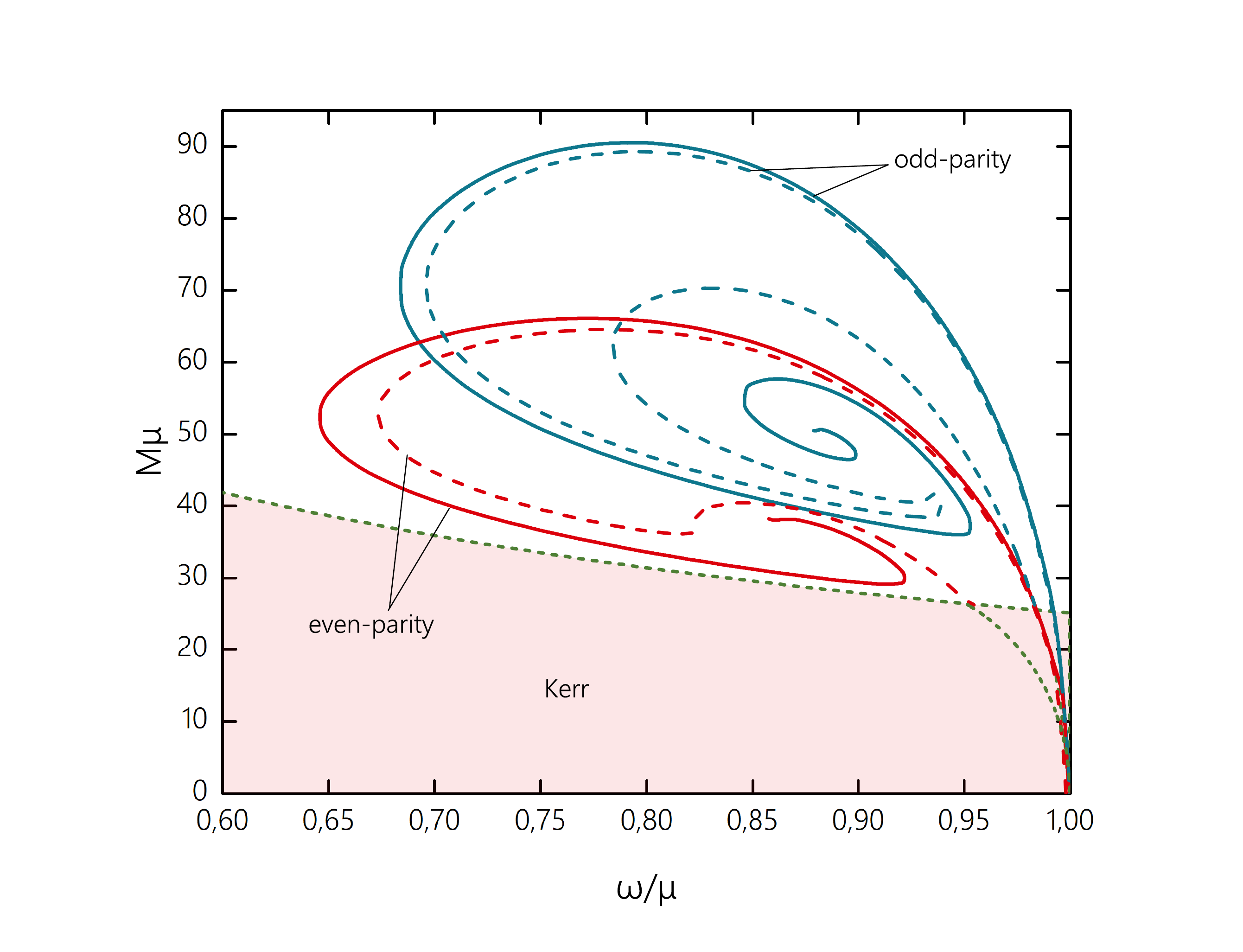}
    \end{center}
    \caption{\small
Comparison of parity-odd and parity-even $n=1$ boson stars and hairy BHs:
ADM mass $M$ vs frequency $\omega$
for horizon radius parameter $r_h=0$ and $r_h=0.01$.
The shaded area corresponds to the domain of existence of vacuum Kerr BHs.
}
    \lbfig{comparison}
\end{figure}

We have solved the boundary value problem
for the coupled system of nonlinear partial differential equations \re{Einstein}
and \re{scaleq} with boundary conditions \re{bchor}-\re{bcpole} using
a fourth-order finite difference scheme.
The system of equations is discretized on a grid with $101\times 101$ points.
We have introduced the new radial coordinate $x=\frac{r-r_h}{r+c}$,
which maps the semi-infinite region $[0,\infty )$ onto the unit interval $[0,1]$.
Here $c$ is an arbitrary constant which is used to adjust the contraction of the grid.
The emerging system of nonlinear algebraic equations is solved
using a modified Newton method.
The underlying linear system is solved with the Intel MKL PARDISO sparse direct solver.
The errors are on the order of $10^{-9}$.
All calculations have been performed using the CESDSOL
\footnote{Complex Equations -- Simple Domain partial differential equations
SOLver is a C++ library being developed by IP.
It provides tools for the discretization of an arbitrary number
of arbitrary nonlinear equations with arbitrary boundary conditions
on direct product arbitrary dimensional grids
with arbitrary order of accuracy.} library.

Let us now briefly summarize the key results we have obtained
by solving the field equations of the model \re{action}
with the boundary conditions discussed above,
focusing our study on the $n=1$ parity-odd hairy BH solutions.
We recall that such hairy BHs exist only for $n>0$.
The only spherically symmetric solutions with $n=0$ are
globally regular boson stars.

First, we observe that parity-odd $n=1$ solutions exist
in the regular limit $r_h=0$.
They represent a new family of mini boson stars,
which do not possess a flat space limit.
The dependence of these solutions on the frequency $\omega$
is qualitatively similar to that of the spinning boson stars,
that can be linked to flat space Q-balls
because of the presence of higher order self-interaction terms,
and other gravitating solitons
\cite{Volkov:2002aj,Kleihaus:2005me,Kleihaus:2007vk,Radu:2008pp,Kunz:2013wka,Kleihaus:2015iea,Herdeiro:2018djx}.
The existence region of the spinning axially symmetric solutions
is limited by $\omega_{\rm max}=\mu$ from above
and some $\omega_\textrm{min}>0$ from below.

The dependence of the ADM mass and the Noether charge of the parity-odd $n=1$ regular $r_h=0$
configurations on the angular frequency both exhibit a typical inspiralling pattern,
as shown in Fig.~\ref{omega_rh}, left upper plot, for the mass.
As the frequency $\omega$ decreases from $\omega_{\rm max}$, the
mass $M$ gradually increases approaching its maximum at some critical value
of the frequency $\omega_{\rm M}$.
From that point onwards the mass of the solutions on the first fundamental branch decreases
until the minimum frequency  $\omega_\mathrm{min}$ is reached.
At $\omega_\mathrm{min}$ the fundamental forward branch backbends
into a second (backward) branch.
A third (forward) branch and a fourth (backward) branch are also visible
in Fig.~\ref{omega_rh}, upper plot,
forming the first few branches of a spiral.
We note the consistency of this figure with the one presented in
\cite{Wang:2018xhw}, where, however,
no further properties of these hairy BHs were discussed.

In Fig.~\ref{omega_rh}, middle left plot, we show the mass $M$
of the regular parity-odd solutions
versus the total angular momentum $J$.
As seen in the figure, multiple cusps occur in the curves as the branches bifurcate
at the respective maximal and minimal values of the mass.
As expected, for the same set of values of the parameters of the model,
the mass of the parity-odd configurations is higher than the mass
of the parity-even solutions, and the corresponding value of the
minimal frequency  $\omega_\mathrm{min}$ of the parity-odd configurations is higher,
as demonstrated in Fig.~\ref{comparison}.

\begin{figure}[p!]
    \begin{center}
        \includegraphics[width=.4\textwidth, clip = true]{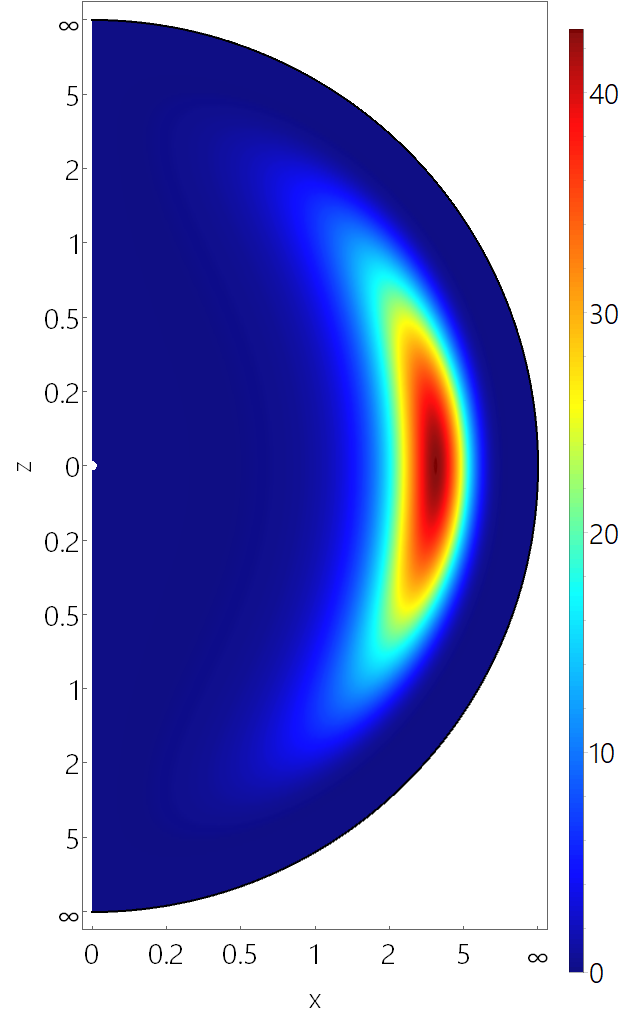}
        \includegraphics[width=.4\textwidth, clip = true]{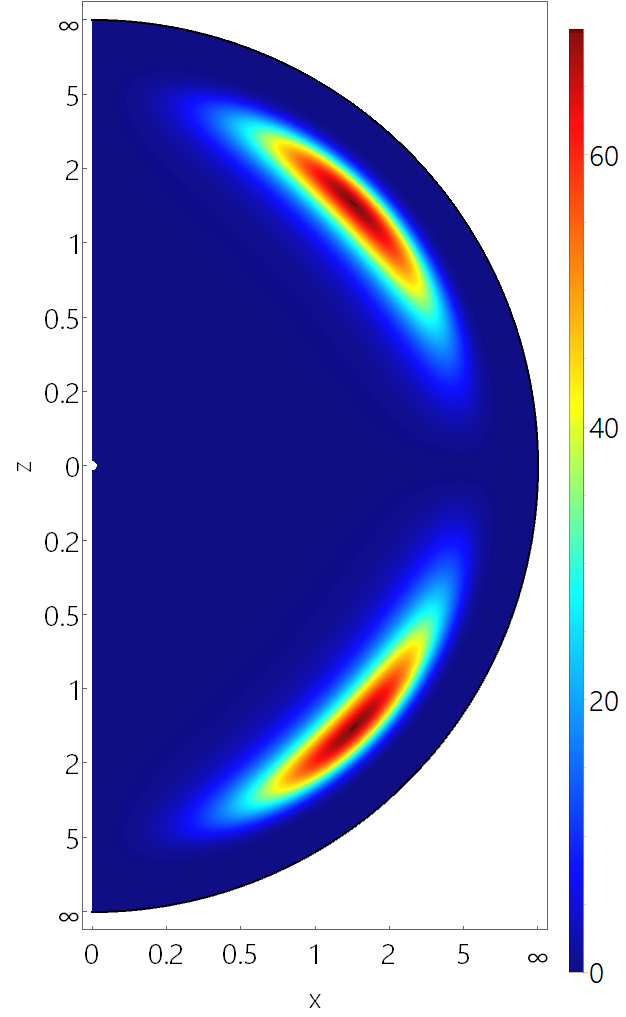}
        \includegraphics[width=.4\textwidth, clip = true]{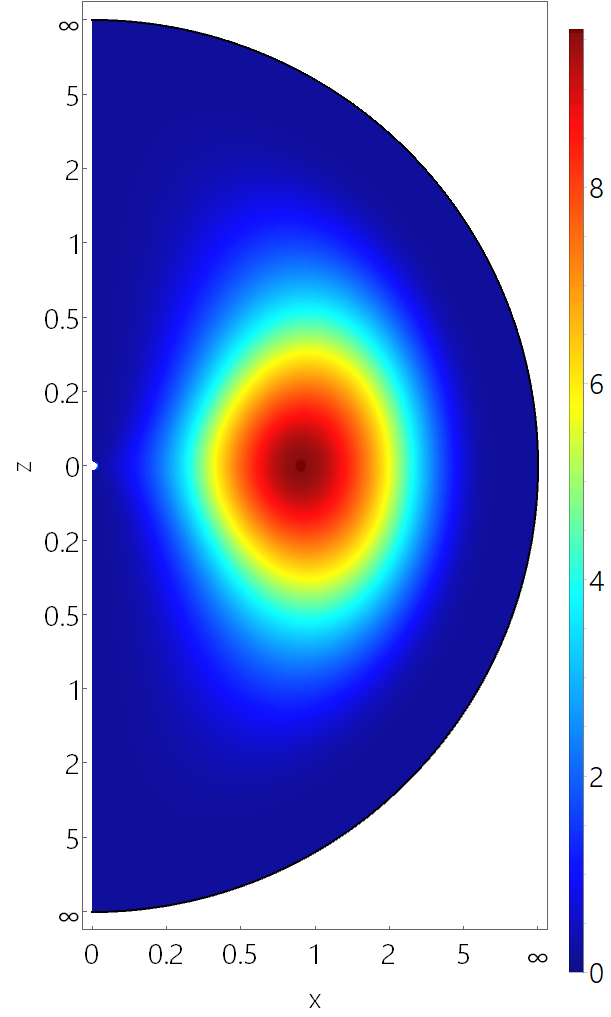}
        \includegraphics[width=.4\textwidth, clip = true]{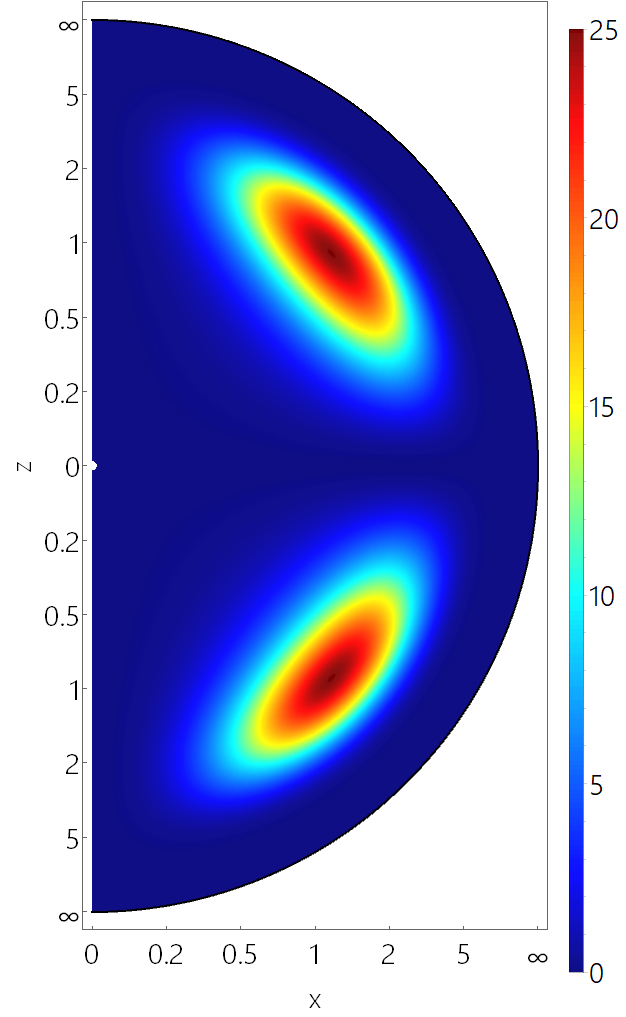}
    \end{center}
    \caption{\small
Stress-energy tensor component $\rho=-T_t^t$ of $n=1$ hairy BHs in the $y=0$ plane
for parity-even (left column) and parity-odd (right column) solutions
with frequency $\omega=0.8$ and horizon radius parameter $r_h=0.01$
on the fundamental forward branch (upper row) and on the backward branch (lower row).}
    \lbfig{endens}
\end{figure}

The domain of existence of parity-odd hairy BHs is scanned by varying the
frequency $\omega$, and the horizon radius parameter $r_h$.
In Fig.~\ref{omega_rh}, upper plot, we illustrate
the dependence of the ADM mass of the spinning hairy BHs on the frequency $\omega$
for a set of values of $r_h$.
As the size of the horizon is increased from zero,
the hairy BHs emerge from the corresponding regular solutions.
When the angular frequency decreases from $\omega_{\rm max}=1$,
the mass of the upper branch solutions increases,
approaching a maximum at some value of frequency $\omega_M$.
As $r_h$ increases, the value of the maximal mass decreases.
The further dependence on the frequency then depends on the value of $r_h$.

For small values of $r_h$ the spiralling type of the critical behaviour
is changed to a multi-branch structure with a smaller number of branches.
The last branch then dives into the
region, where the Kerr BHs reside. Here it merges at some frequency $\omega_{\rm end}<\mu$
with the existence line for the parity-odd scalar clouds around Kerr BHs.
In this limit the scalar field trivializes,
and the branch ends on the respective Kerr BH.
This behaviour is qualitatively similar to the one observed for parity-even
hairy BHs with synchronised hair
\cite{Herdeiro:2014goa,Herdeiro:2015gia,Herdeiro:2015waa}
and also to the one observed in similar models \cite{Herdeiro:2018djx}.

As $r_h$ increases further, the multi-branch structure
is replaced with a two-branch scenario,
with the first (upper, forward) branch
connected to the perturbative excitations at $\omega_{\rm max}=\mu$
and the second (lower, backward) branch, terminating
at the respective Kerr solution as $\omega \to \omega_{\rm end}$.
The minimum angular frequency $\omega_{\rm min}$ increases as $r_h$ increases.
At some point the critical frequency $\omega_{\rm M}$,
which corresponds to the maximal value of the ADM mass,
becomes the minimal allowed frequency $\omega_{\rm min}$.
The maximum value of the frequency along the second branch $\omega_{\rm end}$
is slowly increasing and approaches $\omega_{\rm max}=1$ as the loop shrinks to zero.

Turning to the horizon properties, we note that
the Hawking temperature decreases along the constant $r_h$ curves,
as shown in Fig.~\ref{omega_rh}, middle right plot.
The larger $r_h$, the lower is the temperature on
the fundamental branches of these hairy BHs.
For the smaller values of $r_h$
the value of the scalar field on the horizon $f(r_h)$
increases both along the fundamental forward branch
and along the second backward branch,
while it decreases when the configuration approaches the Kerr limit,
as seen in Fig.~\ref{omega_rh}, lower left plot.
For the higher values of $r_h$,
where the critical frequency $\omega_{\rm M}\sim  \omega_{\rm min}$,
the value of $f(r_h)$ increases along the forward branch,
and decreases along the backward branch.

To illustrate the different scalar field distributions
for the parity-even and parity-odd $n=1$ solutions,
we exhibit in Fig.~\ref{endens} the stress-energy component $\rho=-T_t^t$
for both parity cases for the same parameters $r_h=0.01$ and $\omega=0.8$,
with one set of solutions on the first (forward) branch
and another on the second (backward) branch.
Clearly, for the parity-even solutions the matter distribution
is of the typical toroidal shape,
whereas for the parity-odd solutions it possesses the double-torus structure.
On the second (backward) branch the solutions become more compact.

\begin{figure}[hbt]
    \begin{center}
        \includegraphics[width=.48\textwidth, clip = true, trim = 200 200 200 200 ]{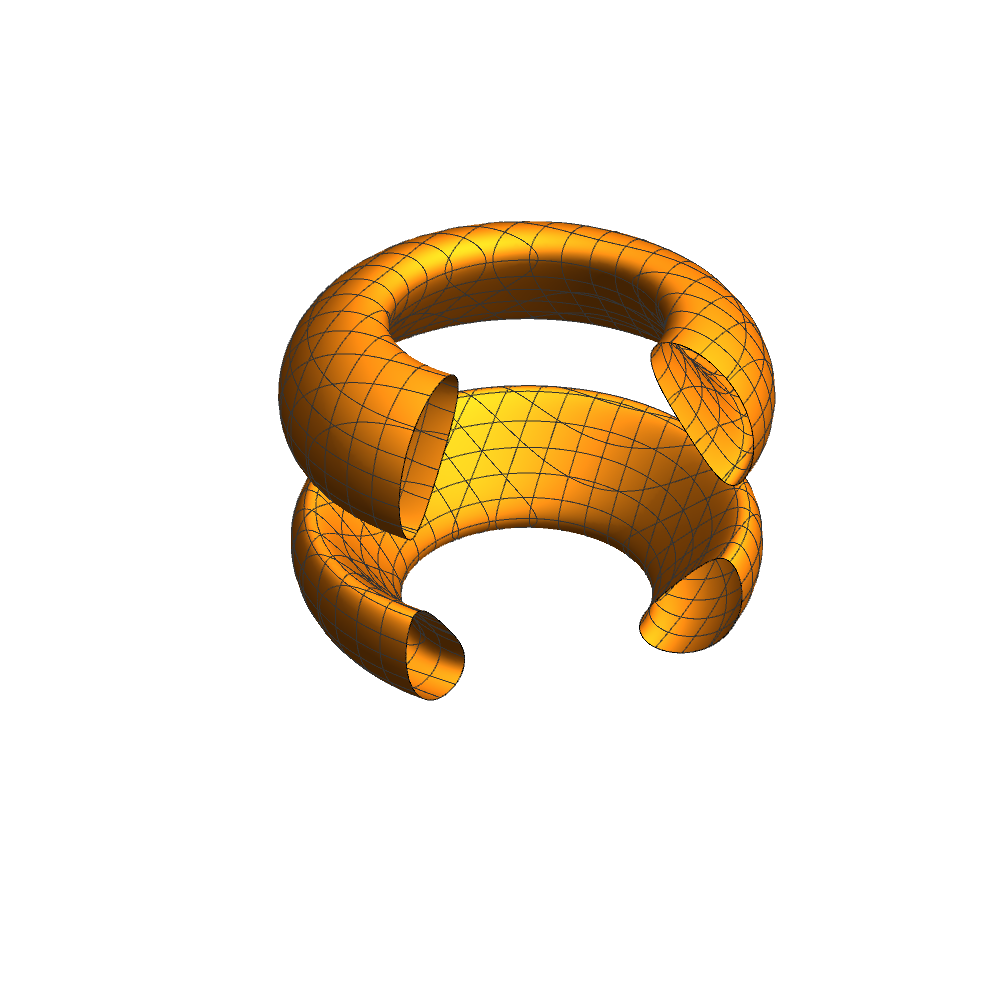}
        \includegraphics[width=.48\textwidth, clip = true, trim = 200 200 200 200]{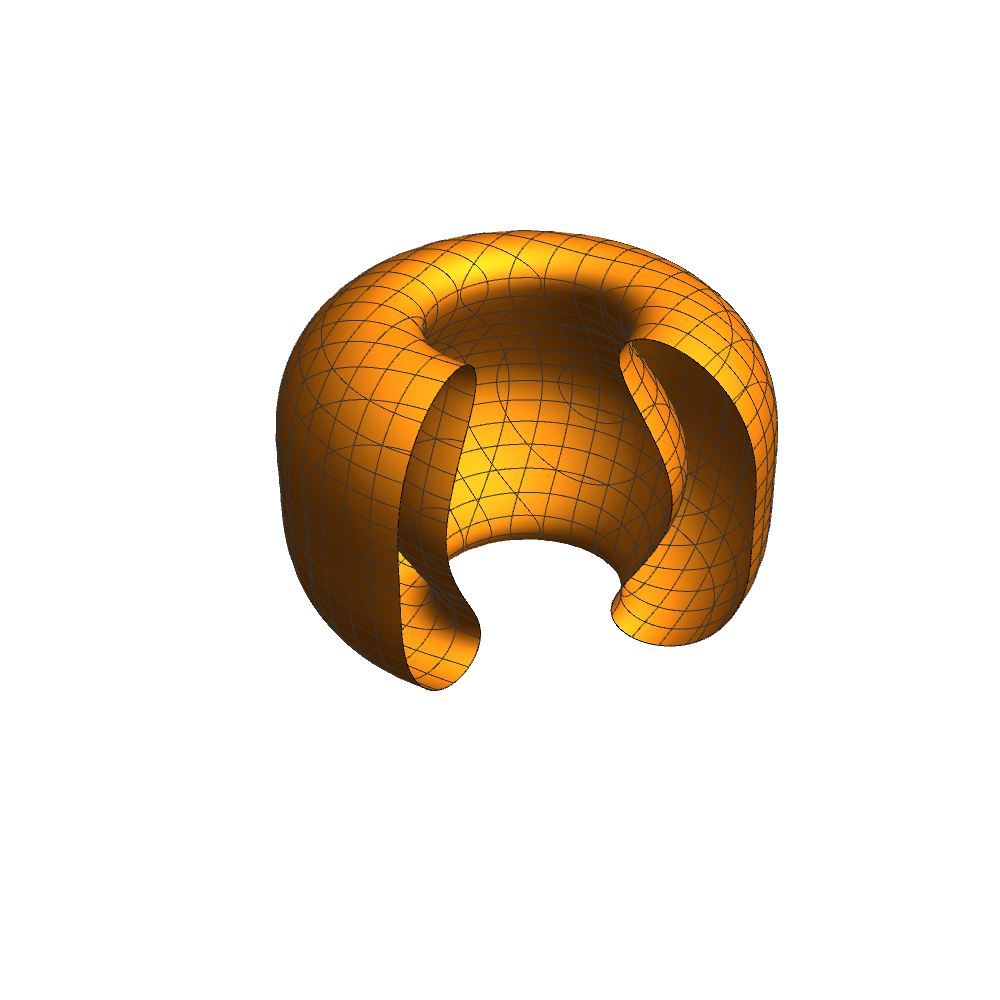}
    \end{center}
    \caption{\small
Ergosurfaces of parity-odd $n=1$ boson stars
with frequency $\omega=0.75$ (left)
and $\omega=0.8$ (right) on the backward branch.}
    \lbfig{ergo_bs}
\end{figure}

\begin{figure}[hbt]
\begin{center}
    \includegraphics[width=.48\textwidth, clip = true, trim = 200 200 200 200]{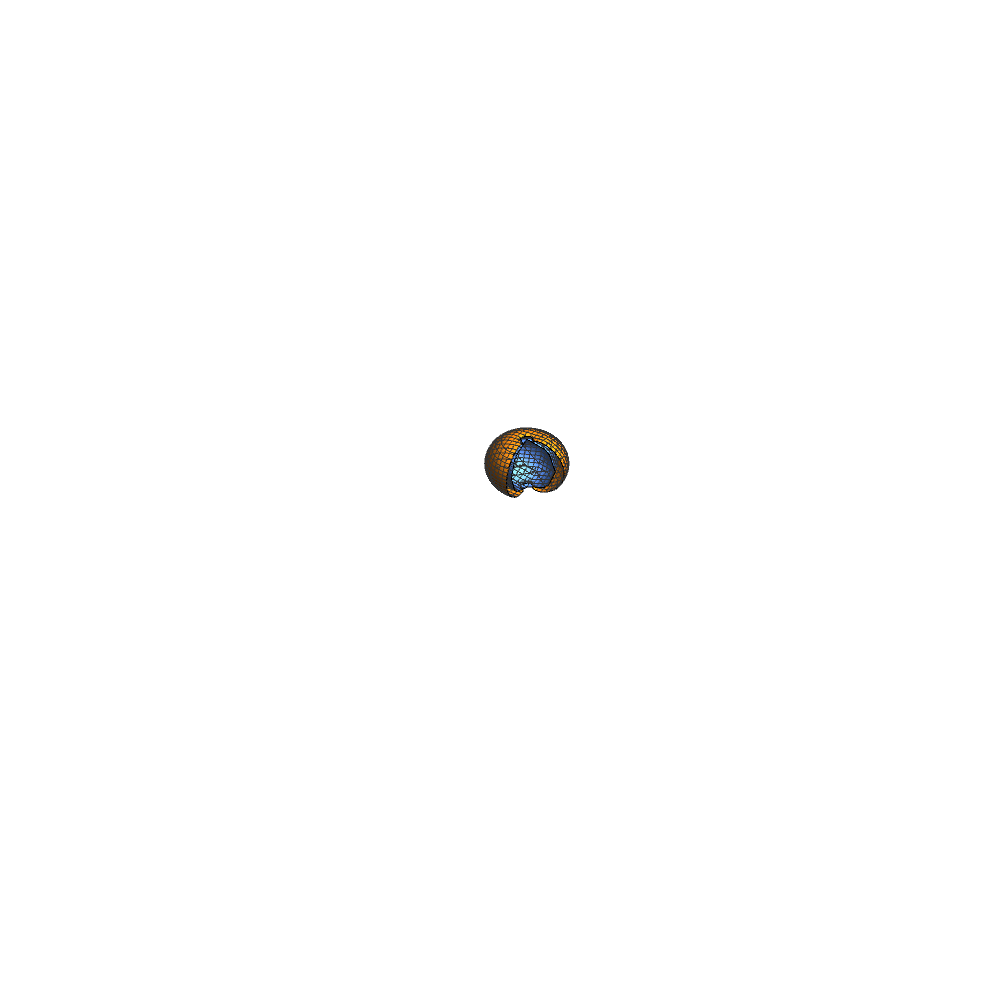}
    \includegraphics[width=.48\textwidth, clip = true, trim = 200 200 200 200]{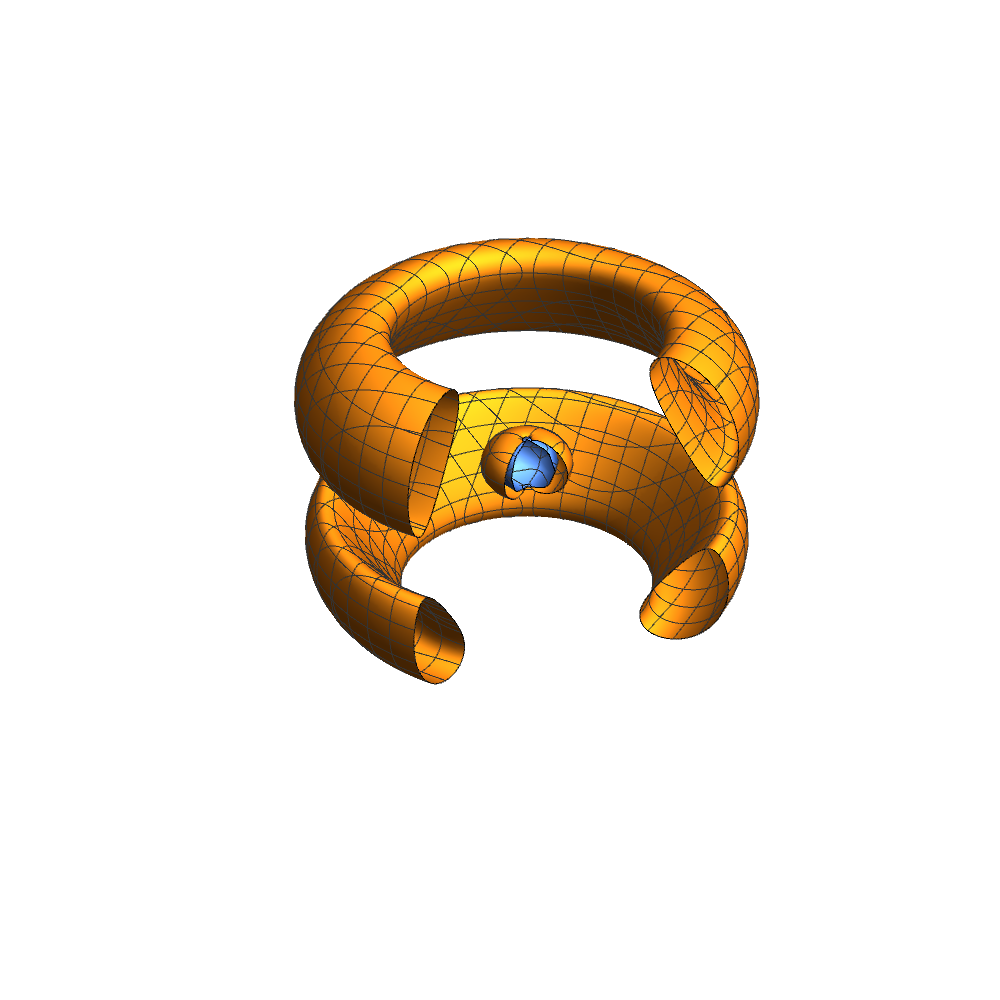}
    \includegraphics[width=.48\textwidth, clip = true, trim = 200 200 200 200]{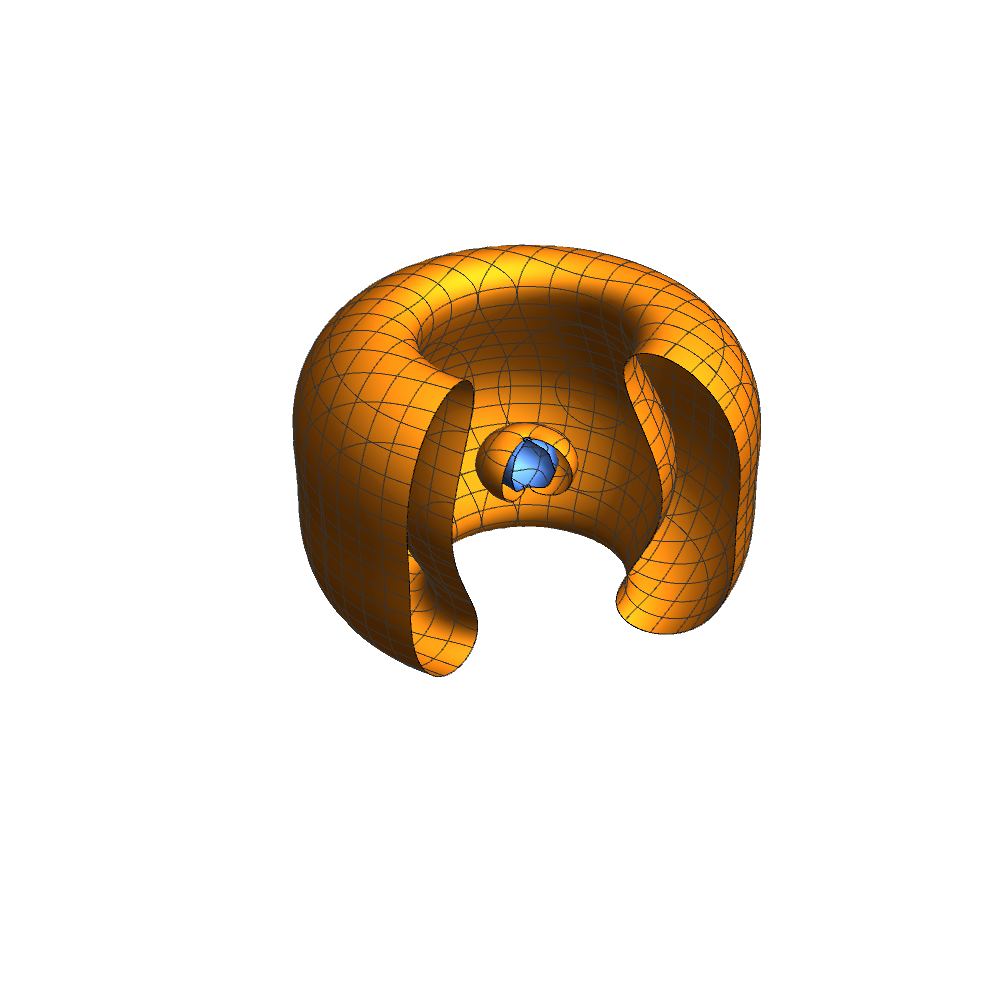}
    \includegraphics[width=.48\textwidth, clip = true, trim = 200 200 200 200]{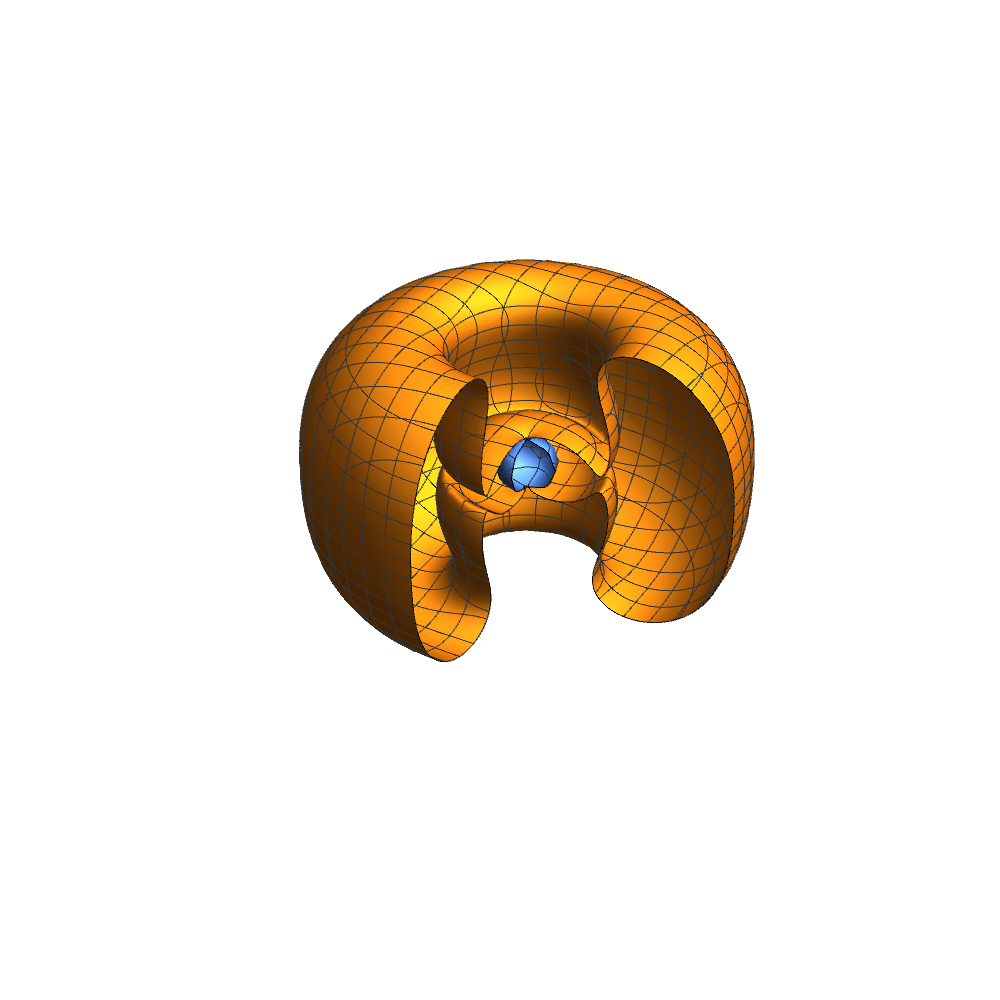}
\end{center}
\caption{\small
Ergosurfaces of parity-odd $n=1$ hairy BHs
with horizon radius parameter $r_h=0.01$
and frequencies $\omega=0.75$ (1), $\omega=0.8$ (2), $\omega=0.85$ (3)
and $\omega=0.93$ (4)
on the backward branch. Blue surfaces represent the horizon.}
\lbfig{ergo_bh}
\end{figure}

Let us next consider the geometry of the ergo-surfaces of the parity-odd spinning BHs,
defined as the zero locus of the normalized time-like Killing vector $\xi \cdot \xi =0$, or
\be
g_{tt}= -F_0 + \sin^2 \theta  F_2 W^2 = 0 \, .
\label{ergo-sf}
\ee
The presence of ergo-regions in rotating boson stars
was realized and investigated in \cite{Cardoso:2007az,Kleihaus:2007vk},
where the non-trivial topology of the ergo-regions was discussed.
For the spinning boson stars ergo-regions appear typically on the fundamental branch
in the vicinity of the maximal mass, where the angular frequency $\omega$
decreases below $\omega_M$ \cite{Kleihaus:2007vk,Herdeiro:2014jaa}.
The topology of the ergo-region of the spinning regular parity-even
solutions is an ergo-torus, $S^1\times S^1$.
Interestingly, for the spinning regular parity-odd
solutions also ergo-double-tori, $(S^1\times S^1) \bigoplus (S^1\times S^1)$,
arise \cite{Kleihaus:2007vk},
as illustrated in Fig.~\ref{ergo_bs}.

The hairy BHs, on the other hand, can either feature a Kerr-like $S^2$ ergo-region,
or an ``ergo-Saturn'', with topology $S^2 \bigoplus (S^1\times S^1)$,
when they possess parity-even scalar hair \cite{Herdeiro:2014jaa}.
For the parity-odd hairy BHs, again Kerr-like ergo-surfaces with $S^2$ topology
are present on the fundamental branch, as shown in Fig.~\ref{ergo_bh}, left plot.
However, on the secondary branches this changes,
and ergo--double-tori appear in addition to the $S^2$ ergo-sphere.
Thus a new type of ergosurface topology arises,
representing an ergo-double-torus-Saturn
$(S^1\times S^1)\bigoplus(S^1\times S^1)\bigoplus S^2$,
illustrated in Fig.~\ref{ergo_bh}, second plot.
Moving on along the branches of constant $r_h$, we observe that
the ergo-double-tori merge into single deformed ergo-tori, and,
as the configurations approach the Kerr limit, into
a single ergo-sphere $S^2$, shown in Fig.~\ref{ergo_bh}, right plot.

\begin{figure}[hbt]
    \begin{center}
        \includegraphics[width=.48\textwidth, trim = 40 20 90 20, clip = true]{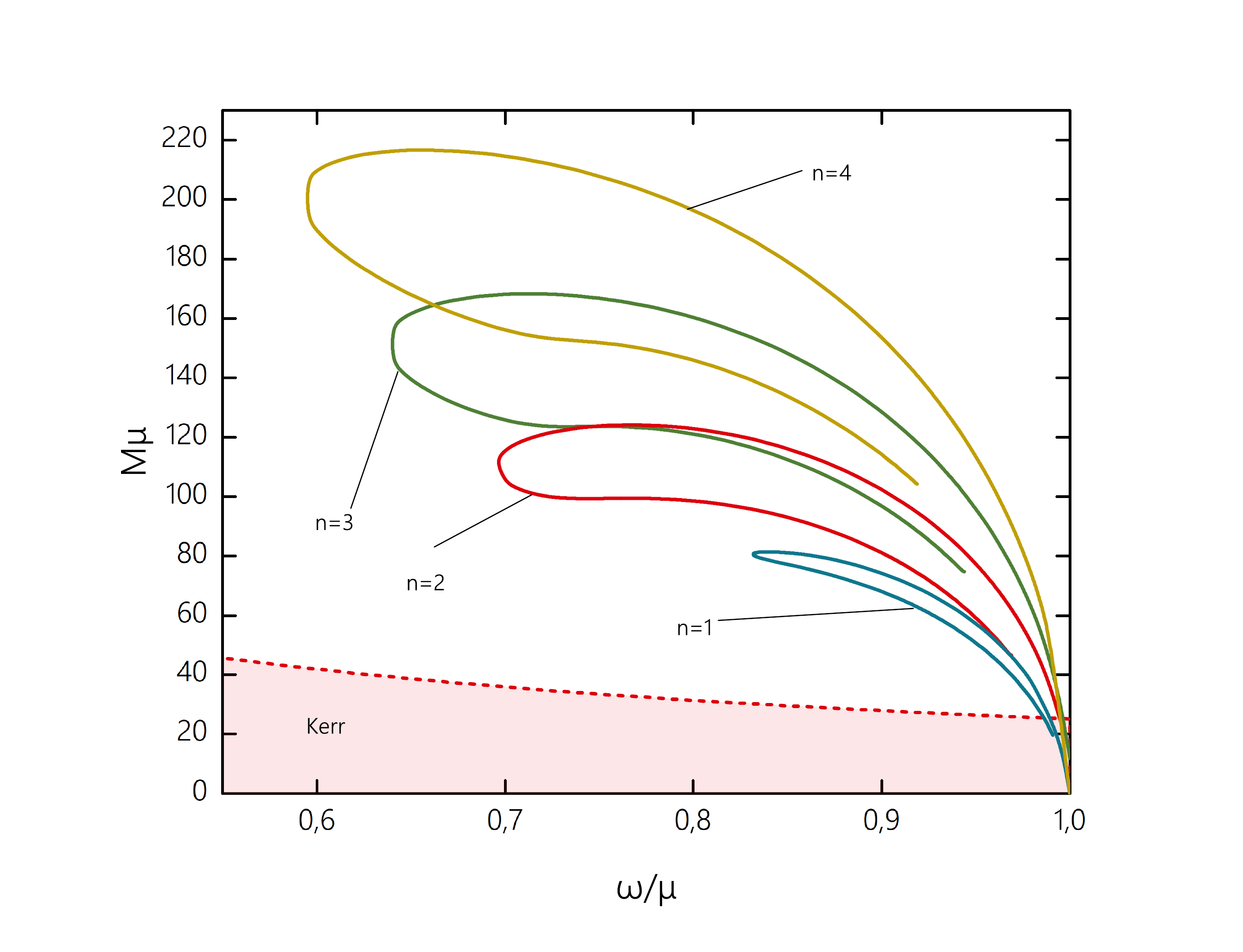}
        \includegraphics[width=.48\textwidth, trim = 40 20 90 20, clip = true]{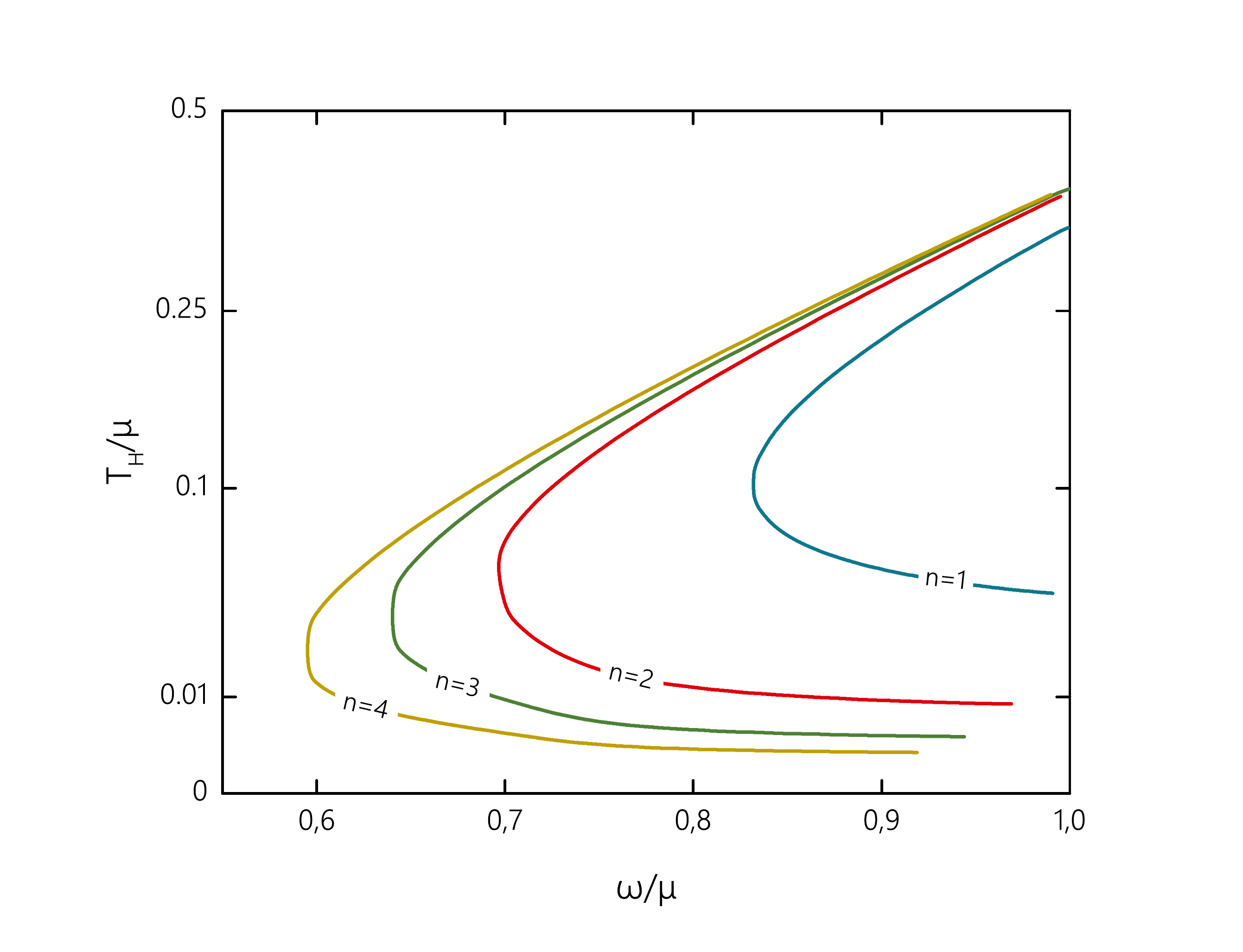}
    \end{center}
    \caption{\small
Properties of parity-odd $n\ge 1$ hairy BHs:
ADM mass $M$ vs frequency $\omega$ (left),
Hawking temperature $T_h$ vs frequency $\omega$ (right),
for the horizon radius parameter $r_h=0.05$.
The shaded area corresponds to the domain of existence of vacuum Kerr BHs.
}
    \lbfig{omega_n}
\end{figure}

We have also studied the parity-odd solutions
with higher values of the winding number $n>1$, shown in Fig.~\ref{omega_n}.
Generally, the branch structure of these solutions is similar
to what we observed for the $n=1$ hairy BHs,
however, their masses are higher and they exist within a larger frequency interval.

\section{Conclusions}

We have considered rotating hairy BHs in Einstein-Klein-Gordon theory,
constructing for the first time BHs with parity-odd scalar hair.
Analogous to their parity-even counterparts, these solutions emerge from
the corresponding parity-odd boson star solutions, when a small finite
event horizon radius is imposed via the boundary conditions.
Their domain of existence is then
determined by the regular boson star solutions on the one hand
and the Kerr BHs on the other hand, where they
merge with the existence line for the parity-odd scalar clouds around Kerr BHs.
In fact, many of the properties of these parity-odd hairy BHs are similar to
those of their parity-even counterparts.

However, the parity-odd scalar field also leads to new features.
Thus the scalar field around the BHs is often concentrated in two tori,
located symmetrically with respect to the equatorial plane.
Consequently the energy density is no longer centered in the equatorial plane.
A further consequence of this distribution of the energy density is
the occurrence of a new type of ergo-surface topology.
Indeed, these rotating hairy BHs may feature
an ergo-double-torus-Saturn
$(S^1\times S^1)\bigoplus(S^1\times S^1)\bigoplus S^2$.

We expect analogous parity-odd hairy BHs to be present in numerous other cases,
where parity-even BHs have been reported. One particular model, where
they arise, is the Friedberg-Lee-Sirling model coupled to Einstein gravity.
These solutions and their properties will be reported elsewhere
\cite{KPS1}.

{\bf Acknowledgements}-- We are grateful to Burkhard Kleihaus and Eugen Radu for inspiring and valuable
discussions. This work was supported in part by the DFG Research Training Group 1620 {\sl Models of Gravity} as
well as by and the COST Action CA16104 {\sl GWverse}. Ya.S. gratefully acknowledges the support of the Alexander
von Humboldt Foundation and from the Ministry of Education and Science of Russian Federation, project No
3.1386.2017. I.P. would like to acknowledge support by the DAAD Ostpartnerschaft Programm.


\end{document}